\documentclass[twocolumn,prl,aps,showpacs]{revtex4}
\usepackage{amssymb}
\usepackage{graphicx}
\usepackage{epstopdf}
\usepackage{amsfonts}
\usepackage{amsmath}
\usepackage{epsfig}
\usepackage{subfigure}
\usepackage{makecell}
\usepackage[colorlinks,anchorcolor=blue,citecolor=blue,linkcolor=black]{hyperref}
\hypersetup{colorlinks=true, citecolor=blue, urlcolor=blue, linkcolor=blue}

\begin{document}

\title{Effect of Stark shift on low-energy interference structure in strong-field ionization}

\author{Weifeng Yang,$^{1,2,3}$}
\email{wfyang@stu.edu.cn}

\author{Jie Li$^{1}$}

\author{Wenbin Jia$^{1}$}

\author{Hongdan Zhang$^{1,2}$}

\author{Xiwang Liu$^{1,2}$}

\author{Ming Zhu$^{1}$}

\author{Xiaohong Song$^{1,2,3}$}

\author{Jing Chen$^{4,5}$}
\email{chen\underline{ }jing@iapcm.ac.cn}

\affiliation{$^{1}$Research Center for Advanced Optics and Photoelectronics, Department of Physics, College of Science, Shantou University, Shantou, Guangdong 515063, China}

\affiliation{$^{2}$ Institute of Mathematics, Shantou University, Shantou, Guangdong 515063, China }

\affiliation{$^{3}$ MOE Key Laboratory of Intelligent Manufacturing Technology, Shantou University, Shantou, Guangdong 515063, China}

\affiliation{$^{4}$ HEDPS, Center for Applied Physics and Technology, Peking University, Beijing 100084, China}

\affiliation{$^{5}$ Institute of Applied Physics and Computational Mathematics, P. O. Box 8009, Beijing 100088,China}

\date{\today}

\begin{abstract}

An improved quantum trajectory Monte Carlo method involving the Stark shift of the initial state, Coulomb potential, and multielectron polarization-induced dipole potential is used to revisit the origin of the low-energy interference structure in the photoelectron momentum distribution of the xenon atom subjected to an intense laser field, and resolve the different contributions of these three effects. In addition to the well-studied radial finger-like interference structure, a ring-like interference structure induced by interference among electron wave packets emitted from multi-cycle time windows of the laser field is found in the low energy part of the photoelectron momentum spectrum. It is attributed to the combined effect of the Coulomb potential and Stark shift. Our finding provides new insight into the imaging of electron dynamics of atoms and molecules with intense laser fields.

\end{abstract}

\maketitle

\section{\label{sec:level1}I. INTRODUCTION}

Photoionization of atoms and molecules under intense laser irradiation is a fundamental process in light--matter interaction \cite{Agostini1979,Becker2002,Gong2017,Song2018}.
Because the intensity of the laser field is very high
and comparable to the potential fields, it dominates the motion of the electrons released from the atoms and molecules.
Therefore, some important issues in
strong-field physics, such as above-threshold
ionization (ATI), high-harmonic generation, and non-sequential double ionization, were described using a semi-classical model \cite{Corkum1993,Keldysh1965,Faisal1973,Reiss1980,Lewenstein1994} in which the ion potential effect on the emitted photoelectron in continuum was ignored and the photoelectron was assumed to be a classical particle. However,  numerous experimental and theoretical evidences have pointed out the faultiness of the semi-classical model due to neglecting the Coulomb potential effect. For example, the low energy structure (LES), i.e., a series of low energy peaks along the laser polarization direction, has been observed in the photoelectron spectra of atoms subjected to an intense infrared laser field \cite{Blaga2009,Quan2009,Wu2012}. The LES was absent in
the semi-classical strong-field approximation simulation without considering the Coulomb potential effect, but it was predicted by the numerical solution of the time-dependent Schr$\ddot{\textrm{o}}$dinger equation (TDSE) \cite{Blaga2009}. Theoretical studies based on Coulomb-corrected semi-classical theories have demonstrated that the LES is related to the long-range Coulomb potential effect \cite{Quan2009,Wu2012,Yan2010,Liu2010,Kastner2012,Guo2013}.

Subsequently, a radial interference structure in the low-energy region of the photoelectron momentum distribution (PEMD), i.e., the so-called fanlike structure, has been widely observed in photoionization of different atoms (He, Ne, Ar, Kr and Xe) \cite{Rudenko2004,Gopal2009, Liu2012}, but the origin of the structure remains obscure and debated \cite{Arbo2006,Arbo2008,Gopal2009,Lai2017,Liu2012}. The Coulomb potential effect on the classical angular momentum distribution and the minimum number of absorbed photons needed to reach the threshold in multiphoton ionization were discussed with a classical trajectory Monte Carlo method \cite{Arbo2006, Arbo2008}. Recently, the PEMD for a hydrogen atom ionized by an intense laser field was investigated using a semi-classical two-step model, which found that the low-energy interference structure (LEIS) is related to the Coulomb potential effect \cite{Shilovski2016}.

Besides the effects of the Coulomb potential, the imprints of multielectron (ME) polarization effect have been found in photoelectron spectra \cite{Pfeiffer2012,Dimitrovski2012,Maurer2014,Madsen2015,Stapelfeldt2015,Kang2018,Le2018,Shilovski2018,Martiny2010,Boulanger2005,Brabec2007}. Shvetsov-Shilovski \emph{et al.} found that the ME polarization effects affect both the tunneling exit point and the subsequent dynamics, which thus decide the photoelectron momentum distribution (PEMD) in elliptically polarized pulses in a method based on tunnel ionization in parabolic coordinates with induced dipole and Stark shift (TIPIS)\cite{Dimitrovski2012}. Further investigation revealed that the ME polarization effect increases the distance from the parent ion to the tunnel exit point of the photoelectron and weakens the interaction between them, so that the photoelectron angular distributions are different for different atomic species \cite{Wang2017}. Recently, it was shown that the relative yields of the LES are enhanced owing to the ME polarization potential on the recolliding electrons \cite{Kang2018}. Moreover, it was found that the electron focusing by the ME polarization-induced dipole potential can induce narrowing of the longitudinal momentum distribution of photoelectrons ionized by a linearly polarized laser pulse \cite{Shilovski2018}.

In contrast to the effects of the Coulomb and ME polarization-induced dipole potential, the Stark shift of the initial state has rarely been noticed, partly because it is usually accompanied by the effect of the ME polarization potential. The Stark shift effect in the strong-field ionization of oriented polar molecules by circularly polarized laser pulses has been investigated using a modified strong-field approximation method \cite{Martiny2010}. In comparison with polar molecules, the Stark shift effect in the ionization of atoms has usually been overlooked. Especially, to the best of our knowledge, the contribution of the Stark shift effect to the interference of the electron wave packet (EWP) is still unclear.

In this work, we revisit the LEIS in the PEMD of the xenon atom, which is a typical ME system, by using an improved quantum trajectory Monte Carlo (IQTMC) method that includes the Stark shift of the initial state, Coulomb potential, and ME polarization potential. By comparing with the TDSE result, we resolve the different effects of these three phenomena on the LEIS, and identify that the ME polarization can enhance the yield of the LEIS, but the detailed pattern of the interference fringes depends sensitively on the Coulomb potential and Stark shift. We find that the Stark shift can substantially affect the phase distribution of the EWP. The semi-classical simulation can reproduce the TDSE result well only when the Stark shift is included.

This paper is organized as follows. In Sec. $\mathbf{II}$, we briefly discuss the IQTMC method that includes the Stark shift of the initial state, Coulomb potential, and polarization potential. In Sec. $\mathbf{III}$, we show the different characteristics of the fringes of the LEIS in PEMDs obtained with different laser intensities. Second, the underlying mechanism of the LEIS is discussed based on the semi-classical
statistical back-trajectory based analysis. The conclusions are presented in Sec. $\mathbf{IV}$.

\section{\label{sec:level2}II. IMPROVED QUANTUM TRAJECTORY MONTE CARLO METHOD}

 In this work, we use the IQTMC method with Feynman's path integral to describe the quantum interference in the linearly polarized laser fields \cite{Kang2018,Shilovski2018,Song2017,Song2016,Lin2016,Yang2016}. The motion of this tunnel-ionized electron is determined by the classical Newtonian equation of motion:
\begin{equation}
\frac{d^{2} \vec{r}}{d t^{2}}=-\vec{F}(t)-\nabla V_{\mathrm{TOT}}(\vec{r}, t),
\end{equation}
where $\vec{F}(t)$ is the electric field of the laser pulse. We consider a linearly polarized laser field $\vec{F}(t)=\vec{F}_{0} f(t) \cos \omega t$ with peak electric field $F_{0}$ and laser frequency $\omega$. The envelope function $f(t)$ is as follows:
\begin{equation}
f(t)=\left\{\begin{array}{ll}{\cos ^{2} \frac{(t-2 T) \pi}{4 T},} & {0 <t \leqslant 2 T}, \\{1,} & {2 T <t \leqslant 6 T}, \\ {\cos ^{2} \frac{(t-6 T) \pi}{4 T},} & {6 T<t \leqslant 8 T}, \\ {0,} & {t>8 T},\end{array}\right.
\end{equation}
where $T$ is the laser optical period.  To be more realistic, the envelope should have turning on and turning off parts. The total ionic potential $V_{\mathrm{TOT}}$, including the Coulomb and ionic core polarization potential, is expressed as
\begin{equation}
V_{\mathrm{TOT}}(\vec{r}, t)=V_{\mathrm{CP}}(\vec{r}, t)+V_{\mathrm{IDP}}(\vec{r}, t),
\end{equation}
where $V_{\mathrm{CP}}(\vec{r})=-Z / |\vec{r}|$ is the Coulomb potential, and $Z=\sqrt{2I_{p}}$ is the effective nuclear charge. The second term denotes the induced dipole potential, i.e., the ionic core polarization induced by the laser field \cite{Dimitrovski2011}, which has the following form:
 \begin{equation}
 V_{\mathrm{IDP}}(\vec{r}, t)=-\alpha_{I} \vec{F}(t) \cdot \vec{r} / r^{3},
 \end{equation}
where $\alpha_{I}$ is the static polarizability of the single charged ion. The ME polarization effect is considered through the induced dipole potential in the above equation. It should be noted that the ME polarization-induced dipole potential is inapplicable at small distances owing to the shielding of the ionic system. Therefore, a cutoff point $r_{c}=\alpha_{I}^{1 / 3}$ is introduced where the core polarization cancels the laser field \cite{Kang2018,Brabec2007}. When $r\leqslant r_{c}$, the electron is nearly field-free and will not experience polarization effects.

In order to solve Eq. (1), we need to obtain the initial position and velocity of the electron. The initial position, i.e., the tunnel exit point can be determined by the one-dimensional Schr$\ddot{\textrm{o}}$dinger equation in a uniform field $F$ in a parabolic coordinate \cite{Dimitrovski2012} as follows:
\begin{equation}
\frac{d^{2} f(\eta)}{d \eta^{2}}+2\left(-\frac{1}{4} I_{p}(F)-V(\eta, F)\right) f(\eta)=0,
\end{equation}
where the effective potential is expressed as
\begin{equation}
V(\eta, F)=-\frac{1 -\sqrt{2 I_{p}(F) / 2}}{2 \eta}-\frac{1}{8} \eta F-\frac{1}{8 \eta^{2}}+\frac{\alpha^{I} F}{\eta^{2}}.
\end{equation}
Physically, Eq. (5) describes a tunneling process for an electron with energy of $-\frac{1}{4} I_{p}(F)$ within an effective potential of $V(\eta, F)$. Therefore, the tunnel exit point $\eta_{e}$ can be determined by solving the equation $V(\eta, F)=-\frac{1}{4} I_{p}(F)$. In Cartesian coordinates, the tunnel exit point is $z_{e} \approx-\eta_{e} / 2$, i.e.,
\begin{equation}
z_{e} \approx-\frac{I_{p}(F)+\sqrt{I_{p}^{2}(F)-4 \beta_{2}(F) F}}{2 F},
\end{equation}
where
\begin{equation}
\beta_{2}(F)=Z-(1+|m|) \sqrt{2 I_{p}(F)} / 2,
\end{equation}
where \emph{m} is the magnetic quantum number.

The Stark shift is included in the IQTMC model by considering the laser field dependent ionization potential \cite{Nakajima2006}.
\begin{equation}
I_{p}(F)=I_{p}(0)+\left(\vec{\mu}_{N}-\vec{\mu}_{I}\right) \cdot \vec{F}+\frac{1}{2}\left(\alpha_{N}-\alpha_{I}\right) F^{2},
\end{equation}
where $I_{p}(0)$ is the field-free ionization potential, $\alpha_{N}$ is the static polarizability of an atom, $\vec{\mu}_{N}$ and $\vec{\mu}_{I}$ are the dipole moments of an atom and its ion, respectively, and \emph{F} is the instantaneous laser field at the tunneling ionization instant of the electron. The field-induced term of Eq. (9) should not exceed $10-20 \%$ of the first term, which introduces an upper bound for the magnitude of the laser intensity \cite{Shilovski2018}.

We assume that the electron starts with zero initial velocity along the direction of the laser field and nonzero initial velocity $v_{\perp}$ in the perpendicular direction. The ionization rate at the tunnel exit point is given by the Ammosov--Delone--Krainov formula \cite{Ammosov1986,Delone1991},
\begin{equation}
\begin{aligned} \Gamma\left(t_{0}, v_{\perp}\right)=\exp \left(-\frac{2\left(2 I_{p}(F)\right)^{3 / 2}}{3 F}\right) \exp \left(-\frac{v_{\perp}^{2} \sqrt{2 I_{p}(F)}}{F}\right) \end{aligned}.
\end{equation}
Therefore, the intensity must not be very low so that the Keldysh parameter $\gamma=\omega Z/F$ is less than or approximately equal to one. Based on the strong-field Feynman's path integral approach \cite{Liu2020,Yang2020}, the phase of the electron trajectory is expressed as
\begin{equation}
\phi_{j}\left(\vec{p}, t_{0}\right)=I_{p}(F)\cdot t_{0}-\int_{t_{0}}^{+\infty}\left\{\vec{v}_{\vec{p}}^{2}(\tau) / 2+V_{\mathrm{TOT}}(\vec{r}, t)\right\} d \tau,
\end{equation}
where $\vec{p}$ is the asymptotic momentum of the $j$th electron trajectory. The probability of each asymptotic momentum is determined as
\begin{equation}
\left|\Psi_{\vec{p}}\right|^{2}=\left|\sum_{j} \sqrt{\Gamma\left(t_{0}, v_{\perp}^{j}\right)} \exp \left(-i \phi_{j}\right)\right|^{2}.
\end{equation}
For the Xe atom, we use the polarizabilities of the statistical theoretical values $\alpha_{\mathrm{Xe}}^{N}$=25.5 a.u., $\alpha_{\mathrm{Xe}}^{I}$=20 a.u. \cite{Shevelko1979}.

\section{\label{sec:level2}III. RESULTS AND DISCUSSION}

\begin{figure}
\includegraphics[width=0.45\textwidth]{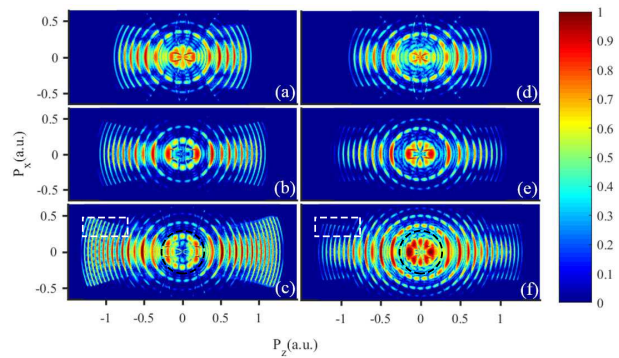}
\caption{Simulated two-dimensional (2D) photoelectron momentum spectra of a Xe atom ionized by a laser pulse with a duration of n = 8 cycles, at a wavelength of 800 nm. (a) and (d) $I=1.05 \times 10^{14} \mathrm{W} / \mathrm{cm}^{2}$ corresponding to the Keldysh parameter $\gamma=0.9829$, (b) and (e) $I=1.5 \times 10^{14} \mathrm{W} / \mathrm{cm}^{2}$, $\gamma=0.8223$, and (c) and (f) $I=2.0 \times 10^{14} \mathrm{W} / \mathrm{cm}^{2}$, $\gamma=0.712$. The left column [panels (a)-(c)]: QTMC calculations considering only the Coulomb potential. The right column [panels (d)-(f)]: IQTMC calculations considering the Coulomb potential, Stark shift, and ME polarization.}
\end{figure}

Figure 1 shows the PEMDs of the Xe atom driven by 800 nm linearly polarized laser fields of different peak intensities $ I = 1.05 \times 10^{14} \mathrm{W} / \mathrm{cm}^{2}$ [$\gamma$ = 0.98, see Figs. 1(a) and 1(d)], $ I = 1.5 \times 10^{14} \mathrm{W} / \mathrm{cm}^{2}$ [$\gamma$ = 0.82, see Figs. 1(b) and 1(e)], and $ I = 2.0 \times 10^{14} \mathrm{W} / \mathrm{cm}^{2}$ [$\gamma$ = 0.71, see Figs. 1(c) and 1(f)]. To resolve the effects of the Coulomb potential, Stark shift, and ME polarization, we compare the PEMDs calculated by the IQTMC method, which considers all the three effects [Figs. 1(d) - 1(f)], with those by the QTMC method, which considers only the effect of the Coulomb potential [Figs. 1(a) - 1(c)]. We find that the Stark shift and ME polarization have obvious influences on the PEMDs, especially on the LEIS. For a low-intensity driven laser field, the LEIS shows radial finger-like fringes for both IQTMC and QTMC simulations \cite{Shilovski2016}. However, the finger-like fringes are shorter in the IQTMC simulation than in the QTMC results [see Figs. 1(a) and 1(d)]. When the laser intensity is increased, the fringes become longer and stronger in the PEDS of the IQTMC simulations [see Figs. 1(e) and 1(f)]. In contrast, the radial finger-like fringes are split by a ring-like destructive interference structure in the PEDS of the QTMC simulations [see Figs. 1(b) and 1(c)]. In the high energy part, the longitudinal fringes at the high P$_{x}$ part of the PEMD are obviously weaker when considering the polarization-induced dipole potential and Stark shift than those without considering these two effects [see the white rectangular region in Figs. 1(c) and 1(f)], demonstrating that the narrowing effect of the longitudinal momentum distributions due to the electron focusing effect of the induced dipole potential can be observed here, which is consistent with that in Mg and Ca \cite{Shilovski2018}. In the following, we mainly focus on the low energy part of the PEMDs ($P_{r}$ $\in$ [-0.27, 0.27] a.u.).

To validate the IQTMC calculations, we perform the TDSE simulation, which can be used as a benchmark \cite{Yang2010,Yang2012,Yang2014}. Figures 2(a) and 2(b) respectively show the LEIS simulated by the TDSE and IQTMC with the same laser parameters as in Figs. 1(c) and 1(f). The IQTMC result obtained by considering all the three effects reproduces well the main feature of the LEIS in the TDSE simulation, i.e., the radial finger-like structure, without splitting. Figs. 2(c)-2(e) show the low-energy parts of the PEMDs by intentionally switching off one or two effects in the semi-classical IQTMC simulations. When only the Coulomb potential is considered, the semi-classical QTMC simulation [see Fig. 2(b)] cannot reproduce the TDSE [see Fig. 2(a)] and the full IQTMC [see Fig. 2(b)] results. As has been stated above, although the radial interference fringes can be discerned, a destructive ring structure splits these fringes. When the ME polarization effect is further added [see Fig. 2(d)], the interference pattern has little change except for a higher probability. Furthermore, when the Stark shift is added in place of the ME polarization effect [see Fig. 2(e)], the destructive ring-like structure moves to the higher energy part, and the radial finger-like structure can now be clearly seen, which reproduces the full IQTMC [see Fig. 2(b)] results well except the lower probability. All these results demonstrate that the LEIS is a combined result of the Coulomb potential and Stark shift, whereas the main effect of ME polarization is the probability enhancement induced by the electron focusing effect of the induced potential \cite{Shilovski2018}.
\begin{figure}
\includegraphics[width=0.45\textwidth]{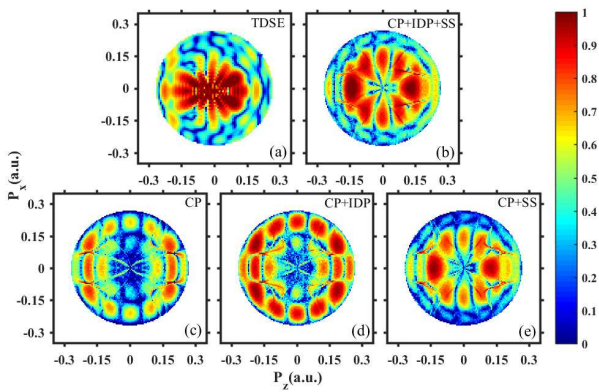}
\caption{LEIS ($P_{r}$ $\in$ [0, 0.27] a.u.) in the 2D photoelectron momentum spectra. (a) Result of the TDSE simulation. (b)-(e) Results of the IQTMC calculations, (b) with the effects of the Coulomb potential, Stark shift, and induced dipole potential, (c) only with Coulomb potential, (d) with inclusion of Coulomb potential and ME polarization, and (e) with inclusion of Coulomb potential and Stark shift. The color scales have been normalized for comparison. The laser parameters are the same as those in Figs. 1(c) and (f). }
\end{figure}

The IQTMC statistical back-trajectory based analysis \cite{Yan2013,Yang2016} can be used to uncover the origin of the ring-like interference structure and demonstrate how the Stark shift affects the ring-like interference of the photoelectron in the PEDMs. We examine the positive half-plane of the PEMDs when considering only the Coulomb potential [see Fig. 3(b)] and when both the Coulomb potential and Stark shift are included [see Fig. 3(c)]. We find that the positive half-plane of the LEIS is mainly contributed by electron trajectories originating from different small time windows of the electric field as shown in Fig. 3(a). For convenient identification, we indicate these time windows with different colors. In the following analysis, if an electron trajectory originates from a certain time window, it would be indicated with the same color [see Fig. 3(d) and 3(e)]. Because the interference structure of the PEMD is related sensitively to the final phase of the electron trajectories, we present in Figs. 3(d) and 3(e) the final phase as a function of P$_{x}$ of the electrons contributing to the central fringes around P$_{z}=0$, i.e., areas denoted by rectangular magenta boxes in Figs. 3(b) and 3(c), respectively. The contribution of scattered electrons is not shown here, as they contribute little to the ring-like interference pattern in this regime.

\begin{figure}
\centering
\includegraphics[width=0.4\textwidth]{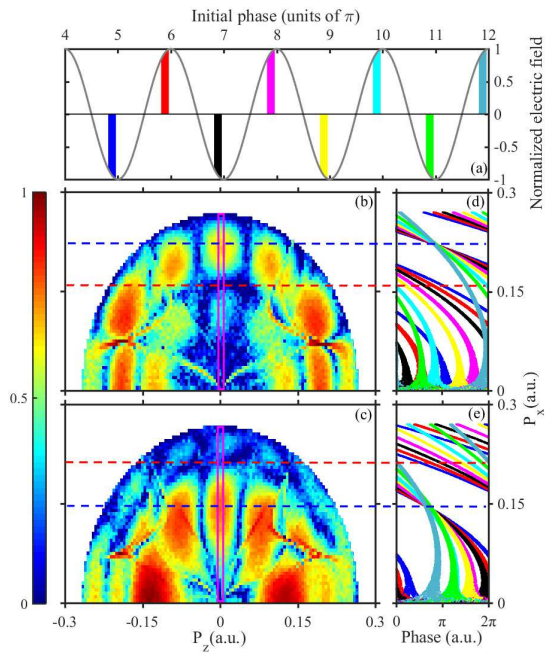}
\caption{(a) Subcycle time windows in which photoelectrons contributing to the positive half-plane of the low energy part of the PEMD originate. (b),(c) positive half planes of the PEDMs shown in Fig. 2(c) and 2(e), respectively. (d), (e) final phase distribution of the electron trajectories emitted from different time windows denoted by the same color in Fig. 3(a) when (d) only the Coulomb potential is considered, and (e) both the Coulomb potential and Stark shift are considered. The blue and red dashed lines denote the phase distributions of constructive and destructive interference, respectively.}
\end{figure}

We find that the final phase of the electron trajectories varies enormously depending on the initial ionization time (indicated by the color) and P$_{x}$. Especially, when only the Coulomb potential is included, the final phase of the electrons from the time windows of different laser cycles are very different in the low P$_{x}$ regime (see stripes of different colors in Fig. 3(d)). It is noted from Eqs. (3) and (11) that the phase related to the Coulomb potential term is hardly dependent on the laser field, which is the main reason for the substantial Coulomb phase difference of the electrons from that of the adjacent laser cycle. When only the Coulomb potential is included, the phase induced by the Coulomb potential term plays an essential role in determining the total final phase of the trajectories. If the final phases of the trajectories from different time windows are widely distributed over the entire phase space ranging from 0-$2\pi$, the electron trajectories emitted from different windows interfere with each other destructively, which induces the interference minimum in the momentum spectrum [indicated by the red dashed line in Fig. 3(b)]. On the contrary, when the final phases of the electron trajectories from different time windows are nearly the same, constructive interference pattern will occur in the PEMD [indicated by the blue dashed line in Fig. 3(d)]. The consistence between the interference maximum and minimum in the PEMD with the phase distribution demonstrates that the ring-like interference structure is actually the result of multicycle interference.

In contrast to the phase related to the Coulomb potential, the phase induced by the Stark shift effect depends strongly on the laser field [see Eqs. (9) and (11)] and changes the total phase of the photoelectron significantly. As a result, the phase difference between the electrons from different cycles is significantly decreased in the low P$_{x}$ regime but increased in the high P$_{x}$ regime [comparing Fig. 3(e) with Fig. 3(d)]. Therefore, the situation is reversed, i.e., destructive interference occurs in the high P$_{x}$ regime [indicated by the red dashed line in Fig. 3(e)] whereas constructive interference (indicated by the blue dashed line) moves to the lower P$_{x}$ part [see Fig. 3(e)] and a clear finger-like LEIS is formed without splitting [see Fig. 3(c)]. Again, the phase distribution analysis is highly consistent with the momentum spectrum. The above findings demonstrate that the Stark shift is a non-negligible effect, which plays an important role in the formation of the LEIS.

\section{\label{sec:level2} IV CONCLUSION}

We theoretically investigated the different contributions of the Coulomb potential, ME polarization, and Stark shift to the 2D-PEDMs of Xe atoms driven by intense infrared laser fields. We especially focused on their influences on the phase of the electron trajectories. We found that the Coulomb potential and Stark shift play an essential role in the final phase of the electron trajectories and therefore have considerable influence on the interference pattern of the LEIS. However, the effect of ME polarization was mainly manifested in the narrowing effect of the longitudinal momentum distribution and the enhancement of the LEIS yield. Our work reveals the importance of including the Stark shift in examining the interference of tunneling EWP. Our findings shed new light on the imaging of ultrafast electron dynamics related to the complex structures of atoms and molecules.

\section{\label{sec:level2} ACKNOWLEDGMENTS }

The work was supported by the National Key Research and
Development Program of China (Grant Nos. 2019YFA0307700
and 2016YFA0401100), NNSF of China (Grant Nos. 11774215, 11674209, 91950101, 11947243, 11334009, and 11425414), Sino-German Mobility
Programme (Grant No. M-0031), Department of Education
of Guangdong Province (Grant No. 2018KCXTD011),
High Level University Projects of the Guangdong Province
(Mathematics, Shantou University), and the Open Fund of
the State Key Laboratory of High Field Laser Physics
(SIOM).


\nocite{*}
\begin{thebibliography}{Starkshift_PRA}

\bibitem{Agostini1979} P. Agostini, F. Fabre, G. Mainfray, G. Petite, and N. K. Rahman, \href{https://journals.aps.org/prl/abstract/10.1103/PhysRevLett.42.1127}{Phys. Rev. Lett.} \textbf{42}, 1127 (1979).

\bibitem{Becker2002} W. Becker, F. Grasbon, R. Kopold, D. B. Milo\v{s}evi\'{o}, G. G. Paulus, and H. Walther, \href{https://www.sciencedirect.com/science/article/pii/S1049250X02800064}{Adv. At Mol. Opt. Phys.} \textbf{48}, 35 (2002).

\bibitem{Gong2017} X. C. Gong, C. Lin, F. He, Q. Y. Song, K. Lin, Q. Y. Ji, W. B. Zhang, J. Y. Ma, P. F. Lu, Y. Q. Liu, H. P. Zeng, W. F. Yang, and J. Wu,  \href{https://journals.aps.org/prl/abstract/10.1103/PhysRevLett.118.143203}{Phys. Rev. Lett.} \textbf{118}, 143203 (2017).

\bibitem{Song2018} X. H. Song, G. L. Shi, G. J. Zhang, J. W. Xu, C. Lin, J. Chen, and W. F. Yang, \href{https://journals.aps.org/prl/abstract/10.1103/PhysRevLett.121.103201}{Phys. Rev. Lett.} \textbf{121}, 103201 (2018).

\bibitem{Corkum1993} P. B. Corkum, \href{https://journals.aps.org/prl/abstract/10.1103/PhysRevLett.71.1994}{Phys. Rev. Lett.} \textbf{71}, 1994 (1993).

\bibitem{Keldysh1965} L. V. Keldysh,  \href{http://www.jetp.ac.ru/cgi-bin/e/index/e/20/5/p1307?a=list}{Sov. Phys. JETP} \textbf{20}, 1307 (1965).

\bibitem{Faisal1973} F. H. M. Faisal,  \href{https://iopscience.iop.org/article/10.1088/0022-3700/6/4/011}{J. Phys. B} \textbf{6}, L89 (1973).

\bibitem{Reiss1980} H. R. Reiss,  \href{https://journals.aps.org/pra/abstract/10.1103/PhysRevA.22.1786}{Phys. Rev. A} \textbf{22}, 1786 (1980).

\bibitem{Lewenstein1994} M. Lewenstein, P. Balcou, M. Yu. Ivanov, A. L. Huillier, and P. B. Corkum,  \href{https://journals.aps.org/pra/abstract/10.1103/PhysRevA.49.2117}{Phys. Rev. A} \textbf{49}, 2117 (1994).

\bibitem{Blaga2009} C. I. Blaga, F. Catoire, P. Colosimo, G. G. Paulus, H. G. Muller, P. Agostini, and L. F. DiMauro,  \href{https://www.nature.com/articles/nphys1228}{Nat. Phys.} \textbf{5}, 335 (2009).

\bibitem{Quan2009} W. Quan, Z. Lin, M. Wu, H. Kang, H. Liu, X. Liu, J. Chen, J. Liu, X. T. He, S. G. Chen, H. Xiong, L. Guo, H. Xu, Y. Fu, Y. Cheng, and Z. Z. Xu,  \href{https://journals.aps.org/prl/abstract/10.1103/PhysRevLett.103.093001}{Phys. Rev. Lett.} \textbf{103}, 093001 (2009).

\bibitem{Wu2012} C. Y. Wu, Y. D. Yang, Y. Q. Liu, Q. H. Gong, M. Wu, X. Liu, X. L. Hao, W. D. Li, X. T. He, and J. Chen,  \href{https://journals.aps.org/prl/abstract/10.1103/PhysRevLett.109.043001}{Phys. Rev. Lett.} \textbf{109}, 043001 (2012).

\bibitem{Yan2010} T. M. Yan, S. V. Popruzhenko, M. J. J. Vrakking, and D. Bauer, \href{https://journals.aps.org/prl/abstract/10.1103/PhysRevLett.105.253002}{Phys. Rev. Lett.} \textbf{105}, 253002 (2010).

\bibitem{Liu2010} C. P. Liu and K. Z. Hatsagortsyan,  \href{https://journals.aps.org/prl/abstract/10.1103/PhysRevLett.105.113003}{Phys. Rev. Lett.} \textbf{105}, 113003 (2010).

\bibitem{Kastner2012} A. K$\ddot{a}$stner, U. Saalmann, and J. M. Rost,  \href{https://journals.aps.org/prl/abstract/10.1103/PhysRevLett.108.033201}{Phys. Rev. Lett.} \textbf{108}, 033201 (2012).


\bibitem{Guo2013} L. Guo, S. S. Han, X. Liu, Y. Cheng, Z. Z. Xu, J. Fan, J. Chen, S. G. Chen, W. Becker, C. I. Blaga, A. D. DiChiara, E. Sistrunk, P. Agostini, and L. F. DiMauro,  \href{https://journals.aps.org/prl/abstract/10.1103/PhysRevLett.110.013001}{Phys. Rev. Lett.} \textbf{110}, 013001 (2013).

 \bibitem{Rudenko2004} A. Rudenko, K. Zrost, C. D. Schr\"{o}ter, V. L. B. de. Jesus, B. Feuerstein, R. Moshammer , and J. Ullrich,  \href{https://iopscience.iop.org/article/10.1088/0953-4075/37/24/L03}{J. Phys. B} \textbf{37}, L407 (2004).

\bibitem{Gopal2009} R. Gopal, K. Simeonidis, R. Moshammer, Th. Ergler, M. D\"{u}rr, M. Kurka, K. U. K\"{u}hnel, S. Tschuch, C. D. Schr\"{o}ter, D. Bauer, J. Ullrich, A. Rudenko, O. Herrwerth, Th. Uphues, M. Schultze, E. Goulielmakis, M. Uiberacker, M. Lezius, and M. F. Kling,  \href{https://journals.aps.org/prl/abstract/10.1103/PhysRevLett.103.053001}{Phys. Rev. Lett.} \textbf{103}, 053001 (2009).

\bibitem{Liu2012} H. Liu, Y. Liu, L. Fu, G. Xin, D. Ye, J. Liu, X. T. He, Y. Yang, X. Liu, Y. Deng, C. Wu, and Q. Gong,  \href{https://journals.aps.org/prl/abstract/10.1103/PhysRevLett.109.093001}{Phys. Rev. Lett.} \textbf{109}, 093001 (2012).

\bibitem{Lai2017} X. Y. Lai, S. G. Yu, Y. Y. Huang, L. Q. Hua, C. Gong, W. Quan, C. Figueira de Morisson Faria and X. J. Liu, \href{https://journals.aps.org/pra/abstract/10.1103/PhysRevA.96.013414}{Phys. Rev. A} \textbf{96}, 013414 (2017).

\bibitem{Arbo2006} D. G. Arb\'{o}, S. Yoshida, E. Persson, K. I. Dimitriou, and J. Burgd\"{o}rfer, \href{https://journals.aps.org/prl/abstract/10.1103/PhysRevLett.96.143003}{Phys. Rev. Lett.} \textbf{96}, 143003 (2006).

\bibitem{Arbo2008} D. G. Arb\'{o}, K. I. Dimitriou, E. Persson, and J. Burgd\"{o}rfer, \href{https://journals.aps.org/pra/abstract/10.1103/PhysRevA.78.013406}{Phys. Rev. A} \textbf{78}, 013406 (2008).

\bibitem{Shilovski2016} N. I. Shvetsov-Shilovski, M. Lein, L. B. Madsen, E. R\"{a}s\"{a}nen, C. Lemell, J.Burgd\"{o}rfer, D. G. Arb\'{o}, and K. T \"{o}k\'{e}si, \href{https://journals.aps.org/pra/abstract/10.1103/PhysRevA.94.013415}{Phys. Rev. A} \textbf{94}, 013415 (2016).


\bibitem{Pfeiffer2012} A. N. Pfeiffer, C. Cirelli, M. Smolarski, D. Dimitrovski, M. Abusamha, L. B. Madsen, and U. Keller,  \href{https://www.nature.com/articles/nphys2125}{Nat. Phys.} \textbf{8}, 76 (2012).

\bibitem{Dimitrovski2012} N. I. Shvetsov-Shilovski, D. Dimitrovski, and L. B. Madsen, \href{https://journals.aps.org/pra/abstract/10.1103/PhysRevA.85.023428}{Phys. Rev. A} \textbf{85}, 023428 (2012).

\bibitem{Maurer2014} D. Dimitrovski, J. Maurer, H. Stapelfeldt, and L. B. Madsen, \href{https://journals.aps.org/prl/abstract/10.1103/PhysRevLett.113.103005}{Phys. Rev. Lett.} \textbf{113}, 103005 (2014).

\bibitem{Madsen2015} D. Dimitrovski and L. B. Madsen,  \href{https://journals.aps.org/pra/abstract/10.1103/PhysRevA.91.033409}{Phys. Rev. A} \textbf{91}, 033409 (2015).

\bibitem{Stapelfeldt2015} D. Dimitrovski, J. Maurer, H. Stapelfeldt, and L. B. Madsen, \href{https://iopscience.iop.org/article/10.1088/0953-4075/48/12/121001}{J. Phys. B} \textbf{48}, 121001 (2015).

\bibitem{Kang2018} H. P. Kang, S. P. Xu, Y. L. Wang, S. G. Yu, X. Y. Zhao, X. L. Hao, X. Y. Lai, T. Pfeifer, X. J. Liu, J. Chen, Y. Cheng, and Z. Z. Xu,  \href{https://iopscience.iop.org/article/10.1088/1361-6455/aabce0}{J. Phys. B} \textbf{51}, 105601 (2018).

\bibitem{Le2018} C. T. Le, V. H. Hoang, L. P. Tran, and V. H. Le,  \href{https://journals.aps.org/pra/abstract/10.1103/PhysRevA.97.043405}{Phys. Rev. A} \textbf{97}, 043405 (2018).

\bibitem{Shilovski2018} N. I. Shvetsov-Shilovski, M. Lein and L. B. Madsen,  \href{https://journals.aps.org/pra/abstract/10.1103/PhysRevA.98.023406}{Phys. Rev. A} \textbf{98}, 023406 (2018).

\bibitem{Martiny2010} D. Dimitrovski, C. P. J. Martiny, and L. B. Madsen,  \href{https://journals.aps.org/pra/abstract/10.1103/PhysRevA.82.053404}{Phys. Rev. A} \textbf{82}, 053404 (2010).

\bibitem{Boulanger2005} T. Brabec, M. C$\hat{o}$t\' e, P. Boulanger, and L. Ramunno, \href{https://journals.aps.org/prl/abstract/10.1103/PhysRevLett.95.073001}{Phys. Rev. Lett.} \textbf{95}, 073001 (2005).

\bibitem{Brabec2007} Z. X. Zhao and T. Brabec,  \href{https://www.tandfonline.com/doi/full/10.1080/09500340601043413}{J. Mod. Opt.} \textbf{54}, 981 (2007).

\bibitem{Wang2017} Y. L. Wang, S. G. Yu, X. Y. Lai, X. J. Liu, and J. Chen, \href{https://journals.aps.org/pra/abstract/10.1103/PhysRevA.95.063406}{Phys. Rev. A} \textbf{95}, 063406 (2017).


\bibitem{Song2017} X. H. Song, J. W. Xu, C. Lin, Z. H. Sheng, P. Liu, X. H. Yu, H. T. Zhang, W. F. Yang, S. L. Hu, J. Chen, S. P. Xu, Y. J. Chen, W. Quan, and X. J. Liu \href{https://journals.aps.org/pra/abstract/10.1103/PhysRevA.95.033426}{Phys. Rev. A} \textbf{95}, 033426 (2017).

\bibitem{Song2016} X. H. Song, C. Lin, Z. H. Sheng, P. Liu, Z. J. Chen, W. F. Yang, S. L. Hu, C. D. Lin, and J. Chen,  \href{https://www.nature.com/articles/srep28392}{Sci. Rep.} \textbf{6}, 28392 (2016).

\bibitem{Lin2016} C. Lin, H. T. Zhang, Z. H. Sheng, X. H. Yu, P. Liu, J. W. Xu, X. H. Song, S. L. Hu, J. Chen, and W. F. Yang,   \href{http://wulixb.iphy.ac.cn/cn/article/doi/10.7498/aps.65.223207}{Acta Physica Sinica} \textbf{65} 223207 (2016).

\bibitem{Yang2016} W. F. Yang, H. T. Zhang, C. Lin, J. W. Xu, Z. H. Sheng, X. H. Song, S. L. Hu and J. Chen,  \href{https://journals.aps.org/pra/abstract/10.1103/PhysRevA.94.043419}{Phys. Rev. A} \textbf{94}, 043419 (2016).

\bibitem{Dimitrovski2011}D. Dimitrovski, M. Abu-samha, L. B. Madsen, F. Filsinger, G. Meijer, J. K\"{u}pper, L. Holmegaard, L. Kalh{\o}j, J. H. Nielsen, and H. Stapelfeldt, \href{https://link.aps.org/doi/10.1103/PhysRevA.83.023405}{Phys. Rev. A} \textbf{83} 023405 (2011)



\bibitem{Nakajima2006} T. Nakajima and G. Buica,  \href{https://journals.aps.org/pra/abstract/10.1103/PhysRevA.74.023411}{Phys. Rev. A} \textbf{74}, 023411 (2006).

\bibitem{Ammosov1986} M. V. Ammosov, N. B. Delone and V. P. Krainov,  \href{http://www.jetp.ac.ru/cgi-bin/e/index/e/64/6/p1191?a=list}{Sov. Phys. JETP} \textbf{64}, 1191 (1986).

\bibitem{Delone1991} N. B. Delone and V. P. Krainov,  \href{https://www.osapublishing.org/josab/abstract.cfm?uri=josab-8-6-1207&origin=search}{J. Opt. Soc. Am. B} \textbf{8}, 1207 (1991).


\bibitem{Liu2020} X. W. Liu, G. J. Zhang, J. Li, G. L. Shi, M. Y. Zhou, B. Q. Huang, Y. J. Tang, X. H. Song, and W. F Yang,  \href{https://journals.aps.org/prl/abstract/10.1103/PhysRevLett.124.113202}{Phys. Rev. Lett.} \textbf{124}, 113202 (2020).

\bibitem{Yang2020} S. D. Yang, X. H. Song, X. W. Liu, H. D. Zhang, G. L. Shi, X. H. Yu, Y. J. Tang, J. Chen, and W. F. Yang,  \href{https://iopscience.iop.org/article/10.1088/1612-202X/aba196}{Las. Phys. Lett.} \textbf{17}, 095301 (2020).


\bibitem{Shevelko1979} V. P. Shevelko and A. V. Vinogradov,  \href{https://iopscience.iop.org/article/10.1088/0031-8949/19/3/010}{Phys. Scr.} \textbf{19}, 275 (1979).

\bibitem{Yang2010} W. F. Yang, X. H. Song, Z. N. Zeng, R. X. Li, and Z. Z. Xu,  \href{https://www.osapublishing.org/oe/abstract.cfm?uri=oe-18-3-2558&origin=search}{Opt. Express} \textbf{18},2558 (2010).

\bibitem{Yang2012} W. F. Yang, X. H. Song, and Z. J. Chen, \href{https://www.osapublishing.org/oe/abstract.cfm?uri=oe-20-11-12067&origin=search}{Opt. Express} \textbf{20}, 12067 (2012).

\bibitem{Yang2014} W. F. Yang, Z. H. Sheng, X. P. Feng, M. L. Wu, Z. J. Chen,and X. H. Song,  \href{https://www.osapublishing.org/oe/abstract.cfm?uri=oe-22-3-2519&origin=search}{Opt. Express} \textbf{22}, 2519 (2014).

\bibitem{Yan2013} T. M. Yan, S. V. Popruzhenko, and D. Bauer,  (\href{https://link.springer.com/chapter/10.1007\%2F978-3-642-35052-8_1}{Springer-Verlag, Heidelberg}, 2013).


\end{thebibliography}
\end{document}